\documentclass[sn-nature]{sn-jnl}

\usepackage{tabularx}
\usepackage{xr} 
\usepackage{rotating}
\usepackage{float}
\usepackage[dvipsnames]{xcolor}
\usepackage{lineno}
\usepackage{hyperref}
\usepackage{enumitem}
\usepackage{graphicx}%
\usepackage{multirow}%
\usepackage{amsmath,amssymb,amsfonts}%
\usepackage{amsthm}%
\usepackage{amsmath}
\usepackage{amssymb}
\usepackage{mathrsfs}%
\usepackage[title]{appendix}%
\usepackage{xcolor}%
\usepackage{textcomp}%
\usepackage{manyfoot}%
\usepackage{booktabs}%
\usepackage{algorithm}%
\usepackage{algorithmicx}%
\usepackage{algpseudocode}%
\usepackage{listings}%
\begin{document}
\title[Article Title]{The Dissipation Theory of Aging: A Quantitative Analysis Using a Cellular Aging Map}


%



\author*[1]{\fnm{Farhan} \sur{Khodaee}}\email{farhank@mit.edu}

\author[1]{\fnm{Rohola} \sur{Zandie}}
\author[1]{\fnm{Yufan} \sur{Xia}}

\author[1,2]{\fnm{Elazer} \sur{R. Edelman}}

\affil*[1]{\orgdiv{Institute for Medical Engineering and Science}, \orgname{Massachusetts Institute of Technology}, \orgaddress{\city{Cambridge}, \postcode{02139}, \state{MA}, \country{USA}}}

\affil[2]{\orgdiv{Department of Medicine (Cardiovascular Medicine)}, \orgname{Brigham and Women’s Hospital}, \orgaddress{\city{Boston}, \postcode{02115}, \state{MA}, \country{USA}}}


\abstract{We propose a new theory for aging based on dynamical systems and provide a data-driven computational method to quantify the changes at the cellular level. We use ergodic theory to decompose the dynamics of changes during aging and show that aging is fundamentally a dissipative process within biological systems, akin to dynamical systems where dissipation occurs due to non-conservative forces. To quantify the dissipation dynamics, we employ a transformer-based machine learning algorithm to analyze gene expression data, incorporating age as a token to assess how age-related dissipation is reflected in the embedding space. By evaluating the dynamics of gene and age embeddings, we provide a cellular aging map (CAM) and identify patterns indicative of divergence in gene embedding space, nonlinear transitions, and entropy variations during aging for various tissues and cell types. Our results provide a novel perspective on aging as a dissipative process and introduce a computational framework that enables measuring age-related changes with molecular resolution.}




\maketitle

\section{Introduction}


Aging is a universal biological process, yet dissecting its underlying mechanisms remains an elusive challenge in biology. Numerous theories have sought to explain the aging process, ranging from the programmed aging theory \cite{weismann1891essays,longo2005programmed}, phenoptosis \cite{skulachev1997aging}, evolutionary theories \cite{medawar1952unsolved, williams1957pleiotropy}, and damage accumulation \cite{harman1955aging,kirkwood1977evolution,gladyshev2016aging}. Meanwhile, some scientists have shifted focus from overarching theories to cataloging the observable hallmarks of aging, revealing the complexity and diversity of aging mechanisms across living systems \cite{lopez2013hallmarks}. While these approaches provide important insights, they often limit their investigations only to biological observations \cite{kirkwood2005understanding} and therefore, a unifying framework to understand aging remains absent.

Aging is not merely the passage of time but reflects the dynamic state of the system, encompassing its molecular, cellular, and also phenotypic changes over time. In this sense, aging has to be studied as a dynamic process. Biological organisms are essentially a form of dynamical systems composed of numerous interacting components that operate across multiple scales \cite{bertalanffy1934theoretische}. These components follow complicated rules and feedback loops to maintain homeostasis and functional integrity. Here, we propose to use dynamical systems theories to study the changes during aging. Specifically, we will use ergodic theory and Hopf decomposition theorem to analyze the system \cite{hopf1937ergodentheorie}. According to this theory, any dynamical system can be decomposed into two distinct components: the conservative part which is recurrent, and therefore in the phase space (the space of all possible states of the system), where almost every point cycles between the same states, and the dissipative part which is non-recurrent and points escape over time and do not return to initial state \cite{walters2000introduction}. We propose that in biological dynamical systems (BDS), aging represents the predominance of dissipative forces which leads to deviation from recurrent states and increased entropy over time. However, quantifying dissipation in dynamical systems is challenging and requires accurate formalism of the system at hand. 

The rise of data-driven approaches has emerged as novel methods to approximate the behavior of complex dynamical systems. Traditional methods for studying biological dynamical systems often rely on simplified models that fail to account for the inherent nonlinearity and high dimensionality of complex processes. Recent advancements in computational techniques, especially in machine learning and data-driven modeling, have made it possible to reconstruct these systems computationally  \cite{schmidt2009distilling,brunton2016discovering,legaard2021constructing}. These data-driven techniques have particular relevance for biological systems, where the complexity of interactions—ranging from molecular to organismal scales—makes traditional mechanistic modeling difficult. We have shown previously that language models trained on large-scale multimodal data can capture the nuanced dynamics of these systems by embedding complex biological states into interpretable representations \cite{khodaee2024multimodal}.

In this paper, we introduce a novel approach for studying aging using data-driven dynamical system modeling. First, we will discuss the mathematical foundations of our theory. Next, we will describe the use of a transformer-based model to track the changes in the states during aging. We will train this model on a massive corpus of data from single-cell RNA sequencing (scRNA-seq) studies, spanning 104 age groups and multiple tissues. Our results present a detailed aging map at cellular resolution, which provides insights into the rate of aging across different tissues and cell types. Specifically, we will show dissipation characteristics in the gene embedding space and identify genes with higher amounts of drift and divergence in their embedding. We further characterize dissipative and conservative genes at different stages of lifespan by calculating entropy. By quantifying these changes, we provide a novel theory and a computational framework for quantifying the aging process, thereby contributing to a deeper understanding of aging.

\section{Results}
\subsection{Modeling and Dataset Overview}
\begin{figure} 
    \centering
    \includegraphics[width=\textwidth]{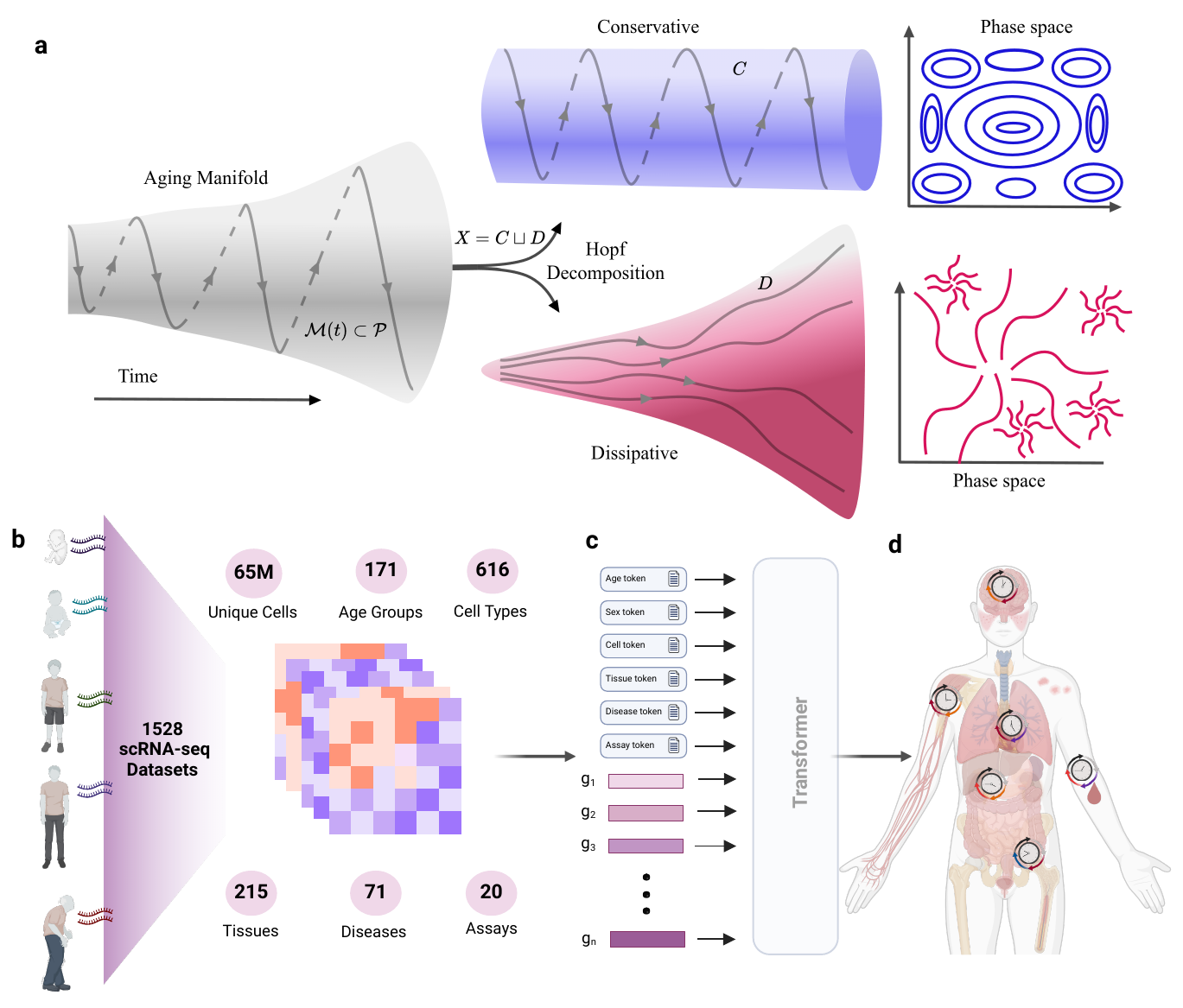}
    \caption{\textbf{Overview of theoretical and computational framework} (a) decomposition of aging manifold in time using Hopf decomposition into conservative and dissipative components (b) aggregation of 1528 single-cell RNA-seq dataset encompassing various age groups, tissues, and cell types, (c) modeling of gene expression using a transformer-based model with metadata information and age token, (d) cellular aging map constructed from the output embedding of the model (Created in https://BioRender.com
)}
    \label{fig1}
\end{figure}

We collected publicly available single-cell RNA sequencing datasets, encompassing over 65 million cells from a wide range of biological sources using the CellxGene platform \cite{megill2021cellxgene}. Specifically, the data included 171 distinct age groups, covering the full spectrum from embryonic stages to old ages, and represented 616 different cell types across 215 tissues, providing a comprehensive view of cellular diversity (Figure \ref{fig1}b). The datasets also span 71 different disease states, allowing us to assess how aging intersects with pathological conditions. The data has been previously standardized and preprocessed to minimize technical batch effects, enhance biological signal consistency, and ensure robustness across different experimental conditions \cite{megill2021cellxgene}. 


We used a multimodal cell foundation model developed previously to map the genotype-phenotype relationship using single-cell transcriptomics data \cite{khodaee2024multimodal}. This model integrates gene expression and metadata information such as chronological age labels with gene expression data (Figure \ref{fig1}c) and creates outputs for each label and gene in the format of an embedding that will be used to construct aging clock at the cellular resolution (Figure \ref{fig1}d) (see the Methods section for more information on modeling \ref{Methods}).   

\subsection{Cellular Aging Map (CAM) Analysis} 
The model's output embeddings are high-dimensional vectors that encapsulate rich information about individual tokens and their contextual relationships within the data. These embeddings are uniquely suitable for capturing tissue-specific and nonlinear aging dynamics. To leverage this capability, we constructed a Cellular Aging Map (CAM) by generating embeddings for each cell type and tissue across distinct phases of aging. CAM provides a detailed landscape of how cellular and molecular features evolve over time.

\subsubsection{Accelerated and Decelerated Aging}
\begin{figure} [!t]
    \centering
    \includegraphics[width=\textwidth]{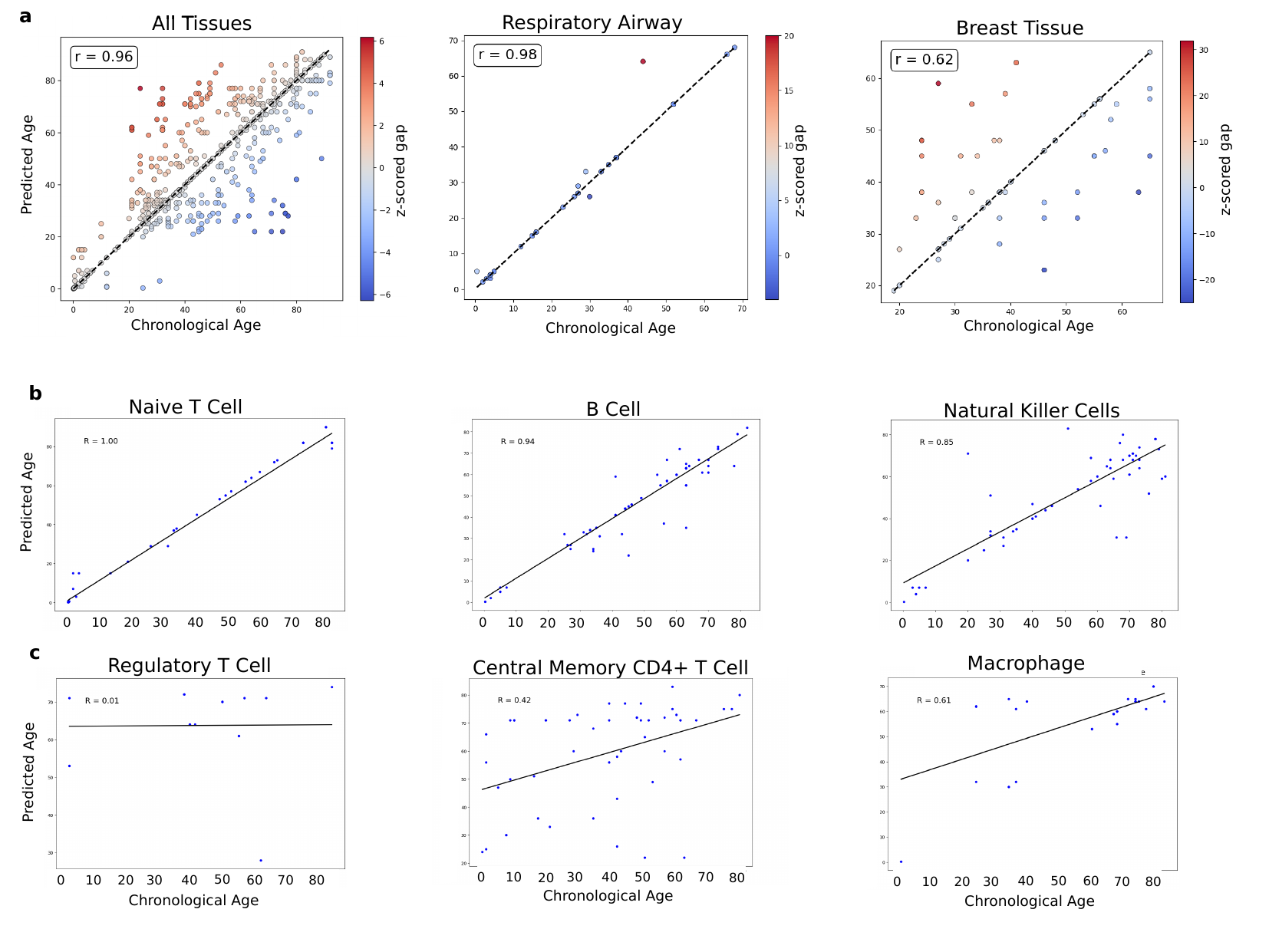}
    \caption{\textbf{Construction of aging clock using age prediction from the model} (a) a 10,000 subsample of all healthy tissues (left), healthy samples from respiratory airway tissue (middle), and healthy samples from breast tissue (right), z-scored gap represents the different between model's prediction and chronological age labels (b) cell-specific age predictions for multiple different cell types with high correlation coefficients that follow chronological age labels, (c) cell types with low correlation coefficient that don't follow the chronological age labels}
    \label{fig2}
\end{figure}

To assess the predictive capabilities of our model, we first tested its ability to estimate the chronological age of samples. Using randomly selected 10000 samples, we evaluated the model's predictions by calculating the z-scored age gap, defined as the difference between the predicted and true ages normalized by the standard deviation. In most cases, the model accurately predicted the cell age, yielding a z-scored age gap close to zero. However, in certain instances, the model predicted significantly higher or lower ages for the samples (Figure \ref{fig2}a). These deviations suggest that the model captures underlying variations in gene expression that reflect processes moving alongside chronological aging and whose effects might be seen as accelerating of decelerating cellular aging. In particular, the distribution of age gaps varied between tissues and cell types, which motivated the development of CAM. 

For instance, predictions for the respiratory airway tissue demonstrated higher accuracy, with a correlation coefficient of r=0.98, and most samples from this tissue were predicted with a z-scored age gap near zero (Figure \ref{fig2}a middle panel). This high degree of concordance suggests a relatively uniform rate of aging within the respiratory airway and possibly stable molecular aging dynamics in this tissue. It might also reflect the rapid turnover of cells in the respiratory tract, an organ system with its own embedded pluripotent stem cells, and one where the time scale of cell turnover is so much faster than the scale of diffusive aging \cite{bowden1983cell}. In contrast, predictions for the breast tissue exhibited higher deviations from the labeled chronological age which shows a heterogeneous rate of aging in this tissue with certain cellular populations exhibiting accelerated or decelerated aging trajectories compared to the expected chronological benchmarks. These observations align with previous findings of tissue-specific aging dynamics \cite{johnson2020systematic, lehallier2020data, argentieri2024proteomic, oh2023organ, tian2023heterogeneous}, and suggest the role of epigenetic phenomena and hormonal cycling, and add more resolution to identify the aging heterogeneity within a tissue.

To further explore aging heterogeneity within tissues, we analyzed the model's age predictions for each cell type individually. This detailed CAM investigation revealed a spectrum of behaviors across different cell populations. Certain cell types, such as Naive T cells, B cells, and natural killer cells exhibited high correlation coefficients between predicted and chronological age (Figure \ref{fig2}b). This could indicate that these cell types experience relatively uniform aging trajectories, with minimal heterogeneity in gene expression patterns over time. On the contrary, other cell types, including regulatory T cells, central memory CD4-positive alpha-beta T cells, and macrophages displayed lower correlation coefficients, indicative of high heterogeneity in their aging profiles (Figure \ref{fig2}b). This variability suggests that our model can differentiate frontline immune cells whose molecular control is systematic and progressive from context-driven cells that are more influenced by environment, exposures, or inflammation, and therefore adding noise to age prediction.  Naive T cells are generated by the thymus at a rate that declines with age, and naive B cells similarly show age-related subtype shifts with immune aging. These cells should therefore track with and even be a marker of chronological age. In contrast, regulatory and central memory T cells and macrophages expand and contract with inflammation, infection, and autounity to maintain a contextual immune balance on a local level and might therefore reflect immune response.

\subsubsection{Age-associated Tissues and Cell Types}

Next, we analyzed the embeddings of tissues and cell types to explore their association with aging. To achieve this, we calculated the similarity score between the embeddings of tissue or cell type tokens and the age token. This similarity score reflects the extent to which a tissue or cell type's molecular profile is influenced by the aging process, as represented by the age token in our model. Higher similarity scores indicate a stronger association with aging, suggesting that the corresponding tissues or cell types are more susceptible to age-related changes. Conversely, lower scores suggest resistance to aging or a slower aging rate based on our dataset. 

\begin{figure}[htbp] 
    \centering
    \includegraphics[width=0.9\textwidth]{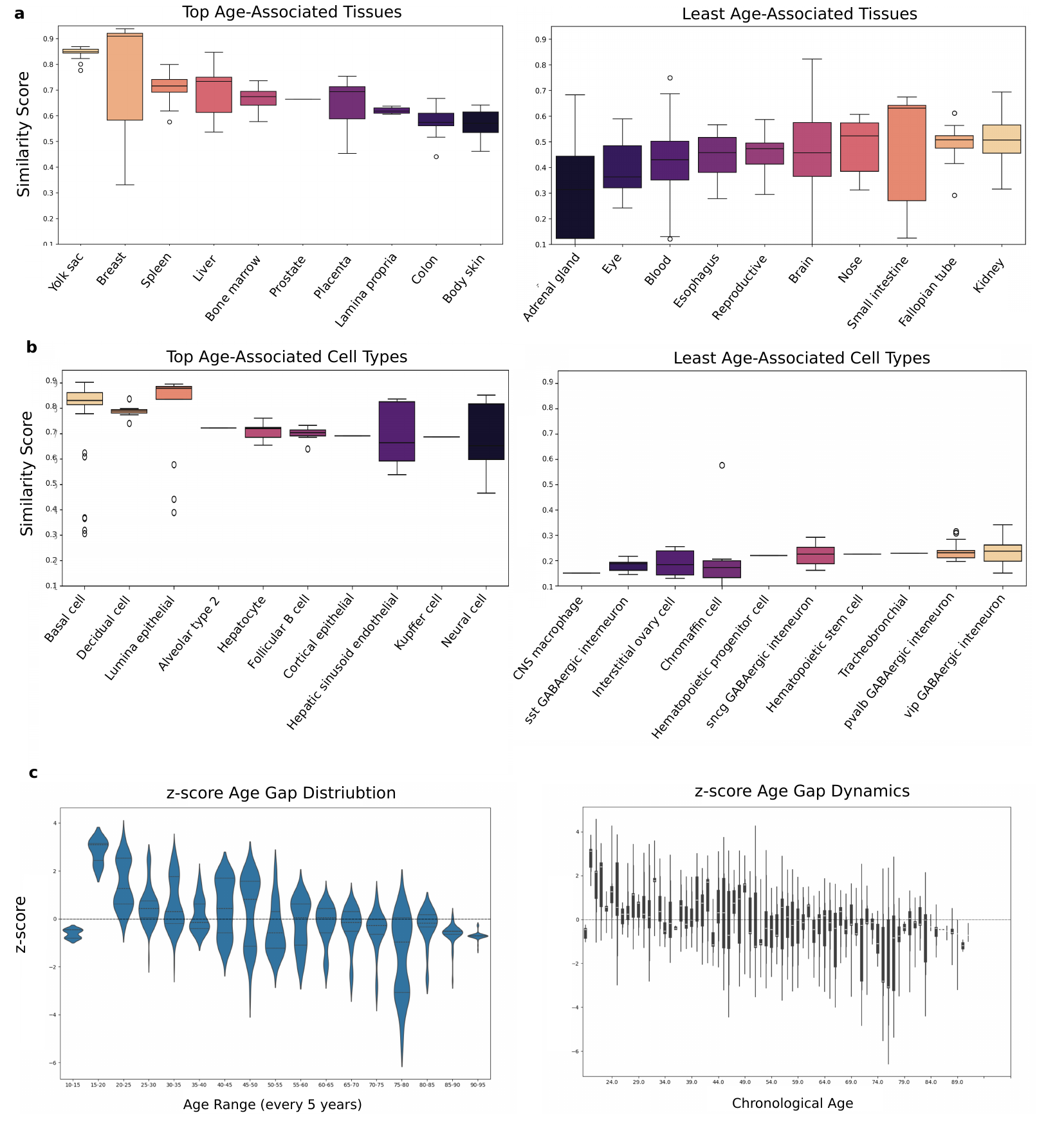}
    \caption{\textbf{Similarity analysis reveals tissues that are more susceptible to or are independent of the aging process} (a) Top age-related tissues that have higher dependency to age token (left) and tissues that are not dependent to age token (right), (b) Top age-related cell types with age token (left), and least age-related cell types, (c) Dynamics of changes in aging using z-scored age gap during different stages of lifespan for age groups every 5 years (left), all samples age labels without grouping (right)}
    \label{fig3}
\end{figure}

To further elucidate the tissue‐ and cell-type-specific patterns of aging, we compared their embeddings against the age token and ranked each tissue or cell type by its similarity score (Figure \ref{fig3}). Boxplots in Figure \ref{fig3}a show the top age‐associated tissues (left) and the least age‐associated tissues (right). Among the highest‐scoring tissues were the yolk sac, breast, spleen, liver, bone marrow, and prostate, all exhibiting elevated similarity to the age token. This suggests that these tissues undergo more pronounced age‐related molecular changes, as they are high-turnover organs in synchrony with systemic aging. By contrast, tissues such as the adrenal gland, eye, blood, and esophagus clustered at the lower end of the similarity spectrum, indicating comparatively reduced susceptibility to—or delayed onset of—molecular aging signals, where there is low turnover and unique aging trajectories

A parallel analysis of cell-type embeddings (Figure \ref{fig3}b) revealed a similar dichotomy. Cell populations, including basal cells, decidual cells, lumina epithelial, hepatocytes, and various immune‐related cell types (e.g., Kupffer cells, cortical epithelial cells, and follicular B cells) ranked among the most age‐associated. Their elevated similarity scores signal a strong coupling between these cell types’ transcriptomic profiles and aging. In contrast, a distinct set of cell types—such as certain interneuron subpopulations and stem/progenitor cells—displayed comparatively lower scores, pointing to more resilient or less dynamically changing expression patterns with advancing age in cells with longer half-lives.

\subsubsection{Nonlinear Dynamics of Age}

To investigate how these age‐associated signatures translate into overall aging trajectories, we computed the age gap for each sample, defined as the difference between the model‐predicted molecular age and chronological age. Figure \ref{fig3}c (left) depicts the distribution of these age gap z‐scores in 5‐year increments across the adult lifespan. Notably, the violin plots demonstrate broad heterogeneity at most ages, potentially due to individual‐level variability in molecular signatures. This could also be due to some tissues and cell types within the same chronological age bracket exhibiting different molecular ages. Overall, the dynamics of z-score show underlying biological or environmental factors that modulate the aging process.

The line plot in Figure \ref{fig3}c (right) captures the dynamics of these z‐score age gaps across increasing chronological age in a higher resolution. Although individual trajectories vary, an overall trend emerges in which the magnitude of the age gap tends to increase in early decades, suggesting that the rate of molecular aging accelerates over initial stages of life. However, in later stages (specifically after 4th decade), the average aging gap becomes negative, which indicates deceleration in the aging process. These results demonstrate the nonlinear nature of the aging process, which is consistent with previous studies \cite{shen2024nonlinear}. This motivates track aging in various decades separately to extract contributing molecular shifts during each stage.


\subsection{Temporal Drift in Age Embedding Space} 
In the Methods section, we demonstrated that embeddings capture rich contextual information about the changes occurring during aging. Unlike gene expression, which is represented as a scalar value of gene activity, embeddings encode dynamic relationships and contextual interactions within the genetic landscape. These embeddings are high-dimensional vectors that incorporate both gene activity, genetic interactions, and also the temporal context in which these changes occur, which makes them particularly well-suited for studying the complexity of aging dynamics.

\begin{figure}[!t] 
    \centering
    \includegraphics[width=0.9\textwidth]{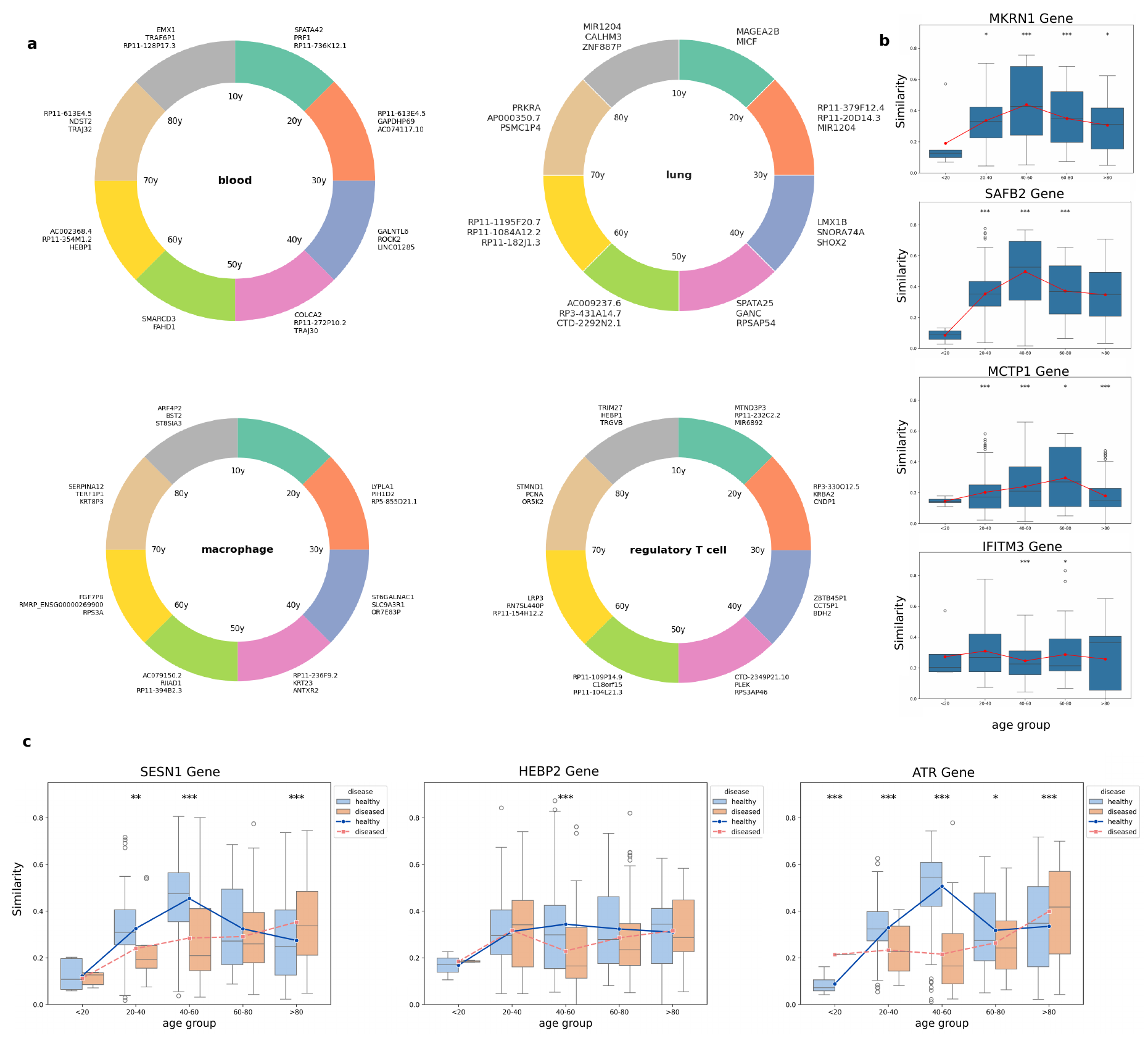}
    \caption{\textbf{Embedding-based analysis reveals temporal and tissue-specific gene dynamics across the aging process.} (a) Cosine similarity between gene embeddings and age embeddings across the lifespan highlights tissue-dependent and cell type–specific aging trajectories. Selected genes are shown for blood and lung tissues, as well as macrophages and regulatory T cells, illustrating stage-specific gene relevance from adolescence to late life. (b) Temporal similarity profiles of representative genes in healthy individuals exhibit diverse aging trajectories. (c) Comparison of gene similarity trajectories between healthy and diseased aging for selected genes shows structured, multi-phase relevance during healthy aging and disruption during disease conditions.}
    \label{fig4}
\end{figure}

One method to capture the temporal dynamics of the aging process is using similarity analysis on embedding vectors. We can calculate cosine similarity among age embeddings to understand the relationship between different age stages, and between age embeddings and gene embeddings to inform the role of genes for each stage. By calculating the cosine similarity between gene embeddings and age embeddings across different stages of the lifespan, we observed distinct variations in the roles that genes play at each stage (Figure \ref{fig4}a-c). Specifically, the temporal similarity trajectories of individual genes suggest that their relevance to the aging process is both tissue and stage-specific.

For example, for blood tissue, genes like \textit{SPATA42} and \textit{PRF1} contribute more to aging during adolescence, but in later stages, for example between 70 to 80 years old, genes such as \textit{TRAJ32} and \textit{NDST2} have higher impact on blood aging. A similar pattern was observed for other tissues such as lung, and also specific cell types, such as macrophages and regulatory T cells (Figure \ref{fig4}a) 

Genes have distinct patterns in their contribution to aging. Some genes exhibit sharply peaked interactions at distinct time points while others maintain broader relevance across multiple age stages. For instance, \textit{MKRN1} and \textit{SAFB2} displayed transient similarity peaks in early to mid-life, suggesting potential regulatory roles during periods of developmental or reproductive maturation (Figure \ref{fig4}b). In contrast, genes such as \textit{MCTP1} showed increasing similarity with later age embeddings, implying potential involvement in aging-associated maintenance or stress-response pathways. Interestingly, some genes, such as \textit{IFITM3}, which is involved in antiviral defense, showed relatively stable similarity across the lifespan, suggesting constitutive or aging-independent activity.  (Figure \ref{fig4}b).

We also compared the gene embedding variations during healthy and diseased aging (Figure \ref{fig4}b). Specifically, \textit{SESN1}, \textit{HEBP2}, and \textit{ATR}, exhibited complex similarity trajectories with multiple local maxima in healthy cases. However, these trajectories were altered in the presence of age-related disease. For example, \textit{ATR} showed diminished association during adulthood in diseased samples, but in old ages showed higher similarity, potentially reflecting impaired DNA repair mechanisms first and then overcompensation. These shifts suggest that aging-associated diseases disrupt the coordinated temporal regulation of key protective genes, which can guide how pathology may override or derail otherwise adaptive aging programs.

\subsection{Aging Is a Dissipative Process}
The CAM analysis demonstrates that chronological time is an inadequate aging metric and does not sufficiently capture the biological aging process, as certain tissues display significant age gaps and the nonlinear nature of age-related changes. To deepen our understanding of the complex aging processes, we turn to concepts from dynamical systems theory, particularly dissipation, to explore the mechanisms underlying aging. Dissipation, in this context, can provide a framework for quantifying the loss of system stability and resilience over time, offering a more holistic perspective on the progression of biological aging. In the next section, we will integrate this theoretical approach with our empirical findings to elucidate how tissues and cells transition through different stages of aging and how this process may differ across biological systems. This integrative approach moves beyond static chronological measures and toward a dynamic and system-level understanding of aging.

To further understand the evolution of the dynamical system manifold in aging, we analyzed gene embeddings in relation to the age token. As outlined in the Methods section, we categorized genes into two classes: dissipative and conservative. Dissipative genes exhibit high temporal drift in the embedding space, reflecting their sensitivity to aging and their dynamically shifting contextual relationships. In contrast, conservative genes demonstrate minimal drift, maintaining stable dynamics over time and suggesting consistent roles in cellular processes throughout the aging process. 



\subsubsection{Dissipative and Conservative Genes}

\begin{figure}[!t]
    \centering
    \includegraphics[width=0.9\textwidth]{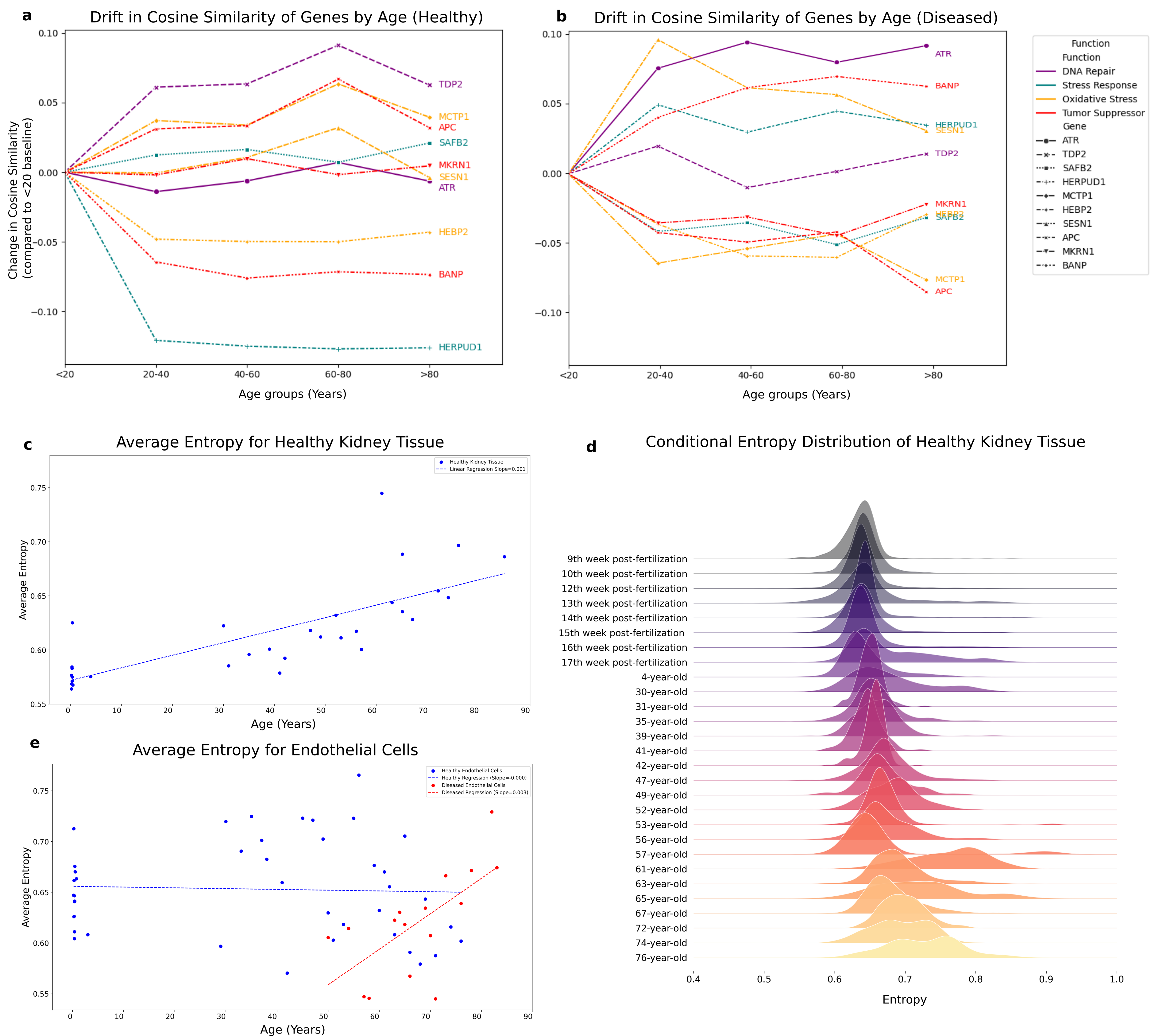}
    \caption{\textbf{Metrics of dissipation for aging in various tissues and cell types.} Extracting conservative and dissipative genes based on temporal drift in gene embedding space during aging in (a) healthy and (b) diseased conditions. Entropy analysis for kidney tissue, (c) average entropy changes and (d)changes in entropy distribution during aging. (e) Endothelial cells entropy changes in aging in healthy and diseased samples}
    \label{fig5}
\end{figure}

Specifically, our analysis revealed distinct examples of these categories. Genes such as \textit{MKRN1}, \textit{SESN1}, and \textit{ATR} were identified as conservative, as their embeddings remained stationary in the embedding space across all stages of life. Conversely, genes such as \textit{ALDH3B1}, \textit{NR2C2}, and \textit{HERPUD1} displayed higher drift from their initial embeddings, categorizing them as dissipative (Figure \ref{fig5}a). Interestingly, the classification of genes as conservative or dissipative appears independent of their biological roles. Both types of genes are involved in a wide range of cellular processes, including DNA repair, stress response, and oxidative metabolism. This observation suggests that conservation and dissipation are universal characteristics intrinsic to the aging process and occur across diverse molecular pathways. This approach allows us to move beyond static measures of gene expression and explore the broader molecular interplay that defines biological aging.

Our analysis revealed alterations in the drift of gene embeddings under diseased conditions, a fundamental shift in gene regulatory dynamics. Specifically, the \textit{ATR} gene, typically characterized as a conservative gene with stable embeddings in healthy cells, exhibited a marked increase in drift within diseased samples (Figure \ref{fig5}b). This shift suggests that \textit{ATR}, which plays a critical role in DNA damage response and cell cycle regulation, may adopt a different expression pattern under pathological conditions. A similar pattern was observed for the \textit{SAFB2} gene, a regulator involved in chromatin organization and stress response, which also displayed increased drift in disease states. These findings could imply a broader trend where genes with stable behavior in healthy tissues become destabilized in the presence of disease. Interestingly, the \textit{TDP2} gene, classified as a dissipative gene due to its high variability in healthy samples, demonstrated an opposite behavior, maintaining a conservative embedding in diseased conditions (Figure \ref{fig5}b). This unexpected stability could suggest a potential compensatory mechanism or a disease-specific regulatory constraint that limits the variability of \textit{TDP2}, possibly to preserve essential functions during cellular stress or damage.

\subsubsection{Entropy Is a Metric of Dissipation and Aging} 
Aging is closely linked to changes in molecular order and structure, with profound implications for the organization of biological systems. From an information theory perspective, it has been proposed that aging is driven by the progressive loss of youthful epigenetic information which essentially leads to increased disorder and reduced fidelity in biological processes \cite{lu2023information}. However, despite this compelling hypothesis, no concrete metrics have been established to quantify this information loss systematically. 

To address this gap, we use entropy as a metric to quantify how aging reshapes the information landscape of gene expression. Entropy, as a measure of uncertainty or disorder in a system, provides a valuable framework for understanding the dynamics of information loss during aging. Specifically, we calculate Shannon entropy based on the model's predictions. For each token in the dataset, we mask the token and task the model to predict it. This process generates a probability distribution over the entire vocabulary, representing the model's uncertainty about the masked token. By applying Shannon entropy to this distribution, we quantify the degree of uncertainty or disorder associated with each prediction. Crucially, we condition these predictions on the age token to capture how the model's uncertainty changes across the aging spectrum (refer to the Method section \ref{method:entropy} for more details). This approach allows us to track the evolution of the information landscape as a function of age. In this context, higher entropy indicates greater uncertainty and suggests a loss of molecular specificity, aligning with the dissipative nature of aging and progressive information degradation.

We performed a detailed entropy analysis across specific tissue and cell types, focusing primarily on kidney tissue and endothelial cells. Our findings reveal a clear trend in healthy kidneys: entropy increases progressively with age, with the rate of increase accelerating during the later stages of life (Figure \ref{fig5}c). Furthermore, the distribution of entropy across different age groups shows a marked rise in entropy variation, which could indicate heightened heterogeneity, potentially signaling either greater uncertainty or increased plasticity in molecular organization within aging kidney tissue (Figure \ref{fig5}d).

In contrast, endothelial cells exhibited a different pattern. No consistent trend of entropy increase was observed across the aging process (Figure \ref{fig5}e). Instead, we identified specific life stages—such as the mid-thirties, early fifties, and early sixties—where entropy exhibited sharp, transient increases. These ``entropy spikes'' were followed by subsequent reductions in entropy, suggesting episodic disruptions in molecular organization that are later compensated or stabilized. This dynamic behavior in endothelial cells highlights a more complex, stage-specific response to aging, distinct from the progressive changes observed in kidney tissue. Interestingly, our analysis showed that diseased endothelial cells consistently exhibited higher entropy compared to their healthy counterparts, demonstrating a higher loss of molecular specificity and an increase in structural or functional disorders (Figure \ref{fig5}e). This heightened entropy suggests that disease conditions exacerbate the underlying molecular chaos within endothelial cells. 

It is fascinating to consider that aging isn’t just a gradual decline. It might involve episodic challenges to molecular stability (entropy spikes), which the body sometimes recovers from. Herein lay a more dynamic view of aging—not just continuous decay propelled by disease, but also adaptive or compensatory recovery in the face of micro-disruptions.

Furthermore, in diseased endothelial cells, entropy demonstrated a clear upward trajectory with aging, which contrasts with the episodic entropy spikes observed in healthy endothelial cells (Figure \ref{fig5}e). This progressive increase suggests that aging and disease act synergistically to amplify molecular uncertainty and disorganization in these cells. The combination of disease-related stress and age-associated decline likely drives this continuous rise in entropy that may underlie the reduced resilience and functional decline observed in diseased endothelial systems over time.

\section{Discussion}
Historically, aging has been examined primarily through a biological lens. Foundational theories-including the  \textit{programmed theory}  \cite{weismann1891essays,longo2005programmed}, \textit{phenoptosis theory} \cite{skulachev1997aging}, \textit{damage accumulation} \cite{medawar1952unsolved}, \textit{antagonistic pleiotropy} \cite{williams1957pleiotropy}, \textit{hyperfunction theory}  \cite{blagosklonny2008aging}, \textit{free radical theory} \cite{harman1955aging, aging2023biomarkers, Harman2009Origin}, \textit{disposable soma theory} \cite{kirkwood1977evolution}, \textit{deleteriome theory}  \cite{gladyshev2016aging}-along with the widely recognized \textit{hallmarks of aging }\cite{lopez2013hallmarks}, have all emphasized biological observations and molecular changes associated with aging \cite{kirkwood2005understanding}.

While these approaches provide important insights, to gain a more comprehensive understanding of aging, it is crucial to integrate perspectives from other disciplines, which can help elucidate the systemic and emergent properties of aging. Raymond Pearl and Ludwig von Bertalanffy followed one such approach. In his book ``Biology of Death'' (1922) \cite{pearl1922biology}, Pearl proposed the ``rate of living'' theory, which related energy expenditure and metabolic rates to lifespan for many organisms. Bertalanffy also proposed a system theory mathematical framework to describe organism's growth over time \cite{von1934untersuchungen}. These theories pioneered the use of more system-level characterization of biological systems in studying aging. 

Our study reconceptualizes aging as a dynamic process and provides a comprehensive exploration of cellular and molecular changes leveraging a novel theoretical framework and advanced modeling techniques. Theoretically, we frame biological organisms as dynamical systems governed by both recurrent (conservative) and non-recurrent (dissipative) forces as described by ergodic theory. We posit that aging emerges from the predominance of dissipative dynamics, leading to deviations from homeostatic, recurrent states and a consequent increase in entropy over time. Because an explicit formulation of aging dynamics is not possible, we provide a novel computational method to learn the dynamics from data implicitly at the cellular level. 

Specifically, we showed that, using our cellular aging map, we generated contextualized embeddings for each gene, tissue, and cell type at each age step. These embeddings enabled capturing both nonlinear and tissue-specific aging trajectories. Model predictions revealed differential rates of aging across tissues and cell types and showed populations with accelerated, decelerated, or resistant aging dynamics. Using embeddings drift analysis, we distinguished between conservative and dissipative genes during both health and disease. To measure the rate of dissipation, we used entropy analysis as an indication of molecular disorder in various tissues. Our results demonstrated entropy as a quantifiable metric of aging-associated information loss. Together, these findings reveal aging as a temporally and spatially heterogeneous process and provide metrics to quantify it at the cellular level. 

Our study has several limitations. One major constraint is our reliance on learning-derived embeddings that, despite capturing complex relationships among genes, tissues, and aging stages, pose challenges for biological interpretability and may obscure direct mechanistic insights. Although complementary analyses such as entropy and temporal drift assessments were employed to enhance understanding, the inherent abstraction of these embedding spaces still limits causal interpretation. Additionally, potential dataset biases, stemming from the underrepresentation of certain tissues, cell types, or age ranges, and a focus primarily on transcriptional dynamics leave gaps in fully capturing rapid or transient aging processes and other layers of the aging landscape, such as epigenetic, proteomic, and metabolomic changes. Addressing these issues with further experimental validation and more comprehensive multi-omics integration will be essential for refining our approach and broadening the applicability of the cellular aging map.

In summary, by expanding on this theoretical and computational framework and refining its applications with other datasets, this work sets the foundation for developing an integrative framework to quantify the loss of stability and resilience during aging, which could lead to more effective therapeutic strategies and personalized medicine applications.

\section{Methods}\label{Methods}
\subsection{Background}

Dynamical systems can be modeled through various approaches depending on whether they evolve continuously or discretely over time. Continuous-time systems are often represented by differential equations, including Ordinary Differential Equations (ODEs) like the motion of a pendulum and Partial Differential Equations (PDEs) such as the heat equation, which describe system behavior in response to time and other variables. For systems that evolve at discrete intervals, difference equations—using linear or nonlinear recursive functions—capture state changes over time, as seen in models of population growth. Dynamical systems can also be described geometrically via phase or state space, where each point represents a system state, and trajectories represent system evolution. Alternatively, vector fields describe time evolution as a flow, mapping system states across time steps, thereby outlining the system’s path on its phase space manifold.

In physics, it's more common to describe the dynamical systems using Hamiltonian or Lagrangian formalisms, which provide a more structured way of capturing the energy dynamics of a system. In systems where randomness or noise plays a role, stochastic differential equations (SDEs) are used. These systems include a random component (usually represented by a Wiener process or Brownian motion) that influences their evolution. Depending on the modeling scheme, we can use different computational algorithms to approximate the future states and overall behavior of the system. Numerical integration of differential equations, Finite Element Method (FEM), Finite Volume Method (FVM), Cellular Automata (CA), Agent-Based Models (ABM) and Stochastic Simulation Algorithms are some of the common algorithms used to approximate the behavior of the system over time.


The study of dynamical systems has traditionally focused on identifying a set of differential equations that describe the time evolution of the system. While this approach works well for simple systems, it becomes increasingly complex as the number of degrees of freedom grows, necessitating new tools and methods of analysis. For instance, coupled systems of differential equations can often be analyzed qualitatively using phase space diagrams, with a focus on the system's stability. Techniques such as examining the tangent space of the phase space with Lyapunov exponents are commonly employed to understand these dynamics and are extremely effective. 

However, when the system has a large number of degrees of freedom, the complexity increases further. In such cases, the presence of multiple positive Lyapunov exponents—often indicative of chaos—shifts the focus from predicting individual trajectories to studying the evolution of the dynamical system \textit{as a whole}. The emphasis moves to understanding the \textbf{manifold} on which the system evolves over time, to capture the overall behavior rather than the specifics of single-particle dynamics. In the context of dynamical systems, a manifold is a mathematical space that locally resembles Euclidean space and serves as a geometric framework to capture the global structure of the dynamics.






These manifolds, commonly referred to as \textit{Lagrangian Coherent Structures (LCS)}, can be generalized to time-dependent manifolds that evolve smoothly over time. To formalize this, consider a phase space denoted by \(\mathcal{P}\), which describes the state space of the system, and a time interval \(\mathcal{T} = [t_0, t_N]\), representing the duration of interest. The dynamics of the system are governed by a time-dependent flow map:

\begin{equation}
    F(t; t_0): \mathcal{P} \to \mathcal{P},
\end{equation}

where \(F(t; t_0)(\mathbf{x}_0)\) represents the position of the trajectory starting at \(\mathbf{x}_0 \in \mathcal{P}\) at the initial time \(t_0\), after evolving under the flow for a time \(t - t_0\). For simplicity, we may assume \(t_0 = 0\), allowing the notation to reduce to: 

\begin{equation}
    F(t): \mathcal{P} \to \mathcal{P}.
\end{equation}

An \textit{invariant manifold} \(\mathcal{M} \subset \mathcal{P}\) satisfies:
\begin{equation}
    F(t; t_0)(\mathcal{M}) = \mathcal{M}, \quad \forall t \in \mathcal{T},
\end{equation}

where every point on \(\mathcal{M}\) remains on the manifold under the flow of the system. This property makes invariant manifolds fundamental in understanding the geometric structure of dynamical systems, as they often represent boundaries between qualitatively different behaviors or act as organizing centers for trajectories.

In a more general setting, we are often interested in the evolution of the manifold itself over extended time periods. To capture this dynamic, we consider a time-dependent family of manifold $\mathcal{M}(t) \subset \mathcal{P}$ that changes smoothly with time. We introduce another flow map $H(t): \mathcal{M}\left(t_0\right) \rightarrow \mathcal{M}(t)$ that captures the intrinsic changes of the manifold:

\begin{equation}
    \mathcal{M}(t)=H\left(t, \mathcal{M}\left(t_0\right)\right)
\end{equation}

where \(H(t)\) maps the manifold at the initial time \(t_0\), denoted by \(\mathcal{M}(t_0)\), to its configuration at a later time \(t\), represented by \(\mathcal{M}(t)\). The evolution of the manifold can then be expressed as:

\begin{equation}
    \mathcal{M}(t) = H(t, \mathcal{M}(t_0)).
\end{equation}

This formulation allows us to distinguish between the intrinsic changes of the manifold and the evolution of the points within the manifold itself. The two flow maps, \(F(t)\), which describes the evolution of individual points in \(\mathcal{P}\), and \(H(t)\), which captures the intrinsic deformation of \(\mathcal{M}\), can be combined to provide a comprehensive description of the dynamics. Specifically, the combined map can be expressed as \(F(t) \circ H(t)\), representing the total dynamical evolution of the manifold. However, in this work, our primary focus is on the map \(H(t)\), which operates on large time scales and governs the long-term intrinsic evolution of the manifold. Understanding \(H(t)\) is key to characterizing the structural dynamics of the system over extended periods.


\subsection{Modeling the Manifold of Gene Expression}
In the case of the dynamics of a gene expressions across a lifetime we are dealing with a dynamical system with many degrees of freedom that creates a changing manifold with time. To model this dynamic we rely on samples taken from different time points from a large population that captures the diversity of this manifold too. 

In this dynamical system, we have the gene expressions and phenotypes associated with them as the total degrees of freedom that change over time. Let $G=(g_1, \cdots, g_n)$ be the set of genes and $Ph=(ph_1, \cdots ,ph_m)$ be the set of phenotypes, then all degrees of freedom are $X=G \cup Ph$. Where \(X\) represents the union of the gene expressions and phenotypes. The evolution of this system over time gives rise to a dynamic manifold \(\mathcal{M}(t) \subset \mathcal{P}\), where \(\mathcal{P}\) is the phase space spanned by \(X\). The structure and evolution of \(\mathcal{M}(t)\) reflect both intrinsic changes in gene expression and their phenotypic consequences.

To capture and model this complex interplay, the flow map \(H(t)\) introduced earlier becomes essential. It represents the evolution of the manifold, incorporating not only temporal dynamics but also variability within the population. By analyzing \(H(t)\), we aim to uncover the long-term dynamics and patterns in gene expression and their associated phenotypic outcomes. To model $H(t, x_1, \cdots, x_n)$, in which $x_i \in X$ and $t \in \mathcal{T}$, we use a Masked Language Model (MLM) approach. More specifically, we are using BERT (Bidirectional Encoder Representations from Transformers) to reconstruct the degrees of freedom as inputs for each time step by randomly masking 85\% of them. Using the MLM the whole dynamics can not be described but it is still able to find the time relationships between the degrees of freedom which is the task of looking at the time evolution of the phase space manifolds.

\subsection{Model Architecture}
The MLM method used here is a powerful tool that can learn contextual relationships among input features in a self-supervised manner. This approach is particularly suited to our problem because gene expressions and phenotypes are inherently interdependent. For example, changes in one gene's expression can cascade into changes in other genes or phenotypes. Moreover,in real-world biological data, we often encounter incomplete datasets with missing gene expressions or phenotypic measurements at certain time points. This mimics the challenges of reconstructing the full manifold \(\mathcal{M}(t)\) from sparse or incomplete observations. Also, this method is scalable for high-dimensional data involved here. We will go over the data and modeling details in the next few sections. 

More specifically, a transformer-based MLM architecture was employed to capture complex interactions between genes and their temporal dynamics. The model used a multi-head self-attention mechanism to learn intricate dependencies between genes. The model was designed to predict masked tokens representing gene expression values, conditioned on the presence of an age token, as well as other biological covariates like tissue type and disease state. This architecture enabled the model to achieve context-aware learning of gene-gene, gene-age, and gene-phenotype relationships, thereby allowing a more holistic understanding of the underlying dynamics of aging. Positional embedding was used to represent each gene uniquely by encoding the positional information relative to other genes within the dataset. This representation allowed the model to capture spatial relationships and dependencies between genes. By incorporating positional embedding, the model could learn not just the gene-specific expression values but also contextualize these values in relation to other genes, thereby improving its ability to capture the underlying structure and interactions within the gene regulatory network.

\subsection{Data Collection}
Our study analyzed over 65 million cells using publicly available single-cell RNA sequencing datasets, representing 616 cell types from 215 tissues across 171 age groups, from embryonic to old age. It also included data on 71 disease states, providing insights into the interplay between aging and pathology. Measured with 20 different assays, the datasets were standardized and preprocessed through normalization, quality control, and batch correction to ensure biological consistency and minimize technical variability \cite{megill2021cellxgene}. 

\subsection{Input Embeddings}
The input embeddings were crafted to integrate multidimensional biological data, combining gene expression profiles with age tokens and other biological metadata to provide a comprehensive representation of the cellular state. Specifically, each embedding was a concatenation of several components; First, the gene expression profiles served as the foundational input, encoding the expression levels of genes in a given cell or tissue. Each profile was normalized and standardized to ensure comparability across samples and minimize technical variability. Next, a discrete representation of the sample's age (labeled as chronological age in the dataset) was incorporated to capture the temporal progression in gene regulation. 

By explicitly embedding age as a feature, the model was conditioned to recognize age-dependent changes in gene expression, enabling it to map temporal trajectories of biological processes. Additional contextual information, such as tissue type and disease state, was included to account for environmental and physiological variations. These metadata were embedded as learnable tokens, allowing the model to disentangle age-related effects from other factors influencing gene expression. Each gene's position within the dataset was encoded using a positional embedding strategy. This approach captured the spatial and regulatory relationships between genes, enabling the model to account for dependencies and interactions within the gene regulatory network. The combined embeddings were fed into the transformer model, where they underwent further transformation through the multi-head self-attention mechanism. 

\subsection{Training Scheme and Evaluation}
The MLM was trained using a masked token prediction task, where portions of gene expression vectors were masked, and the model was tasked with reconstructing them based on contextual information. Model performance was evaluated using reconstruction accuracy and the coherence of generated embeddings across age groups. Embeddings for age tokens, phenotype tokens, and gene tokens were extracted to assess their drift and divergence. In the following section, we will discuss the type of analysis done on these output embeddings.

\subsection{Output Analysis}
While reconstructing genes alone provides insight into the MLM’s predictive ability, our primary interest lies in understanding how the relationships between genes and phenotypes evolve over time. Due to the high dimensionality of the original space, direct visualization and analysis of the dynamical manifold are not feasible. Instead, we focus on the embedding space produced by the MLM, which provides a compact and tractable representation of the manifold.

The key question is whether the embedding space faithfully captures the properties of the dynamical manifold. To address this, we use the BERT embedding function \( E(t, x_1, \cdots, x_n): \mathcal{T} \times X \to \mathbb{R}^d \), where \(d\) is the dimension of the embedding space. Replacing the original flow map \(H\) with the embedding function \(E\), prior research \cite{zhang2022rethinking} shows that BERT embeddings are robust and possess the \textbf{Lipschitz continuity} property:

\begin{equation}
    \left\|E\left(x_1, t_1\right)-E\left(x_2, t_2\right)\right\| \leq L\left(\left\|x_1-x_2\right\|+\left|t_1-t_2\right|\right)
\end{equation}

where \(L\) is the Lipschitz constant. This property implies that small changes in the input state \(x\) or time \(t\) correspond to small changes in the embedding space. 

This robustness is critical because it ensures that the embedding space accurately reflects the dynamics of the original manifold, allowing us to analyze temporal patterns, relationships between genes and phenotypes, and population-wide variations. The Lipschitz continuity guarantees that the embeddings provide a faithful representation of the system's evolution, forming the foundation for the subsequent analysis. Now we can further decompose the embedding space to gain deeper insights into the dynamics of the system. To achieve this, we employ the \textbf{Hopf Decomposition} from ergodic theory, which provides a powerful framework for understanding the behavior of dynamical systems.

\subsubsection{Hopf Decomposition}
In the study of dynamical systems, the \textbf{Hopf Decomposition} offers a way to partition the phase space \(X\) with respect to an invertible and non-singular mapping \(T: X \rightarrow X\). This decomposition divides the phase space into a disjoint union: $X=C \cup D$ in which \(C\) represents the \textit{conservative} part and \(D\) represents the \textit{dissipative} part. These components capture distinct aspects of the system's dynamics:

\begin{itemize}
    \item \textbf{Conservative Component (\(C\))}: In this region of the phase space, the dynamics preserve the underlying structure without permanent deformation. In essence, the measure of subsets remains invariant under the mapping \(T\). Formally, given a measure \(\mu\) of the non-singular dynamical system with the mapping \(T: X \rightarrow X\), the conservative property is defined as:
    \[
    \mu\left(T^{-1}(\sigma)\right) = \mu(\sigma),
    \]
    where \(\sigma \subseteq X\) and \(T^{-1}\) represents the pre-image under the mapping \(T\). This property is also referred to as a \textit{measure-preserving dynamical system}.
    
    \item \textbf{Dissipative Component (\(D\))}: In this region, the dynamics exhibit dissipation, with certain subsets of the phase space gradually ``wandering'' away over time. Specifically, the dissipative part contains a \textit{wandering set}, defined as a subset \(\sigma \subseteq D\) that eventually has no intersection with its earlier states under repeated applications of \(T\). This is expressed mathematically as:
    \[
    \mu\left(T^n(\sigma) \cap \sigma\right) = 0,
    \]
    for \(n \geq 1\), where \(T^n\) represents the \(n\)-fold application of the mapping \(T\).
\end{itemize}

In our modeling of the dynamical system we can write the dynamical embeddings $E(t) \in \mathbb{R}^d$ as conservative and dissipative parts:

\begin{equation}
    E(t)=E_C(t)+E_D(t)
\end{equation}

In this context, $E_C(t)$ and $E_D(t)$ represent the conservative and dissipative components, respectively. It is worth noting that this is just one possible approach to representing the embedding space. However, it is preferred because it highlights the genes as states, making them easier for us to study.

The conservative component should not change much with time $x_{\mathrm{C}}(t) \approx x_{\mathrm{C}}(t+\delta t)$. This can be justified because we shown that the modeling is Lipschitz continuous:
\begin{equation}
    \left\|E_C(t)-E_C(t+\delta t)\right\| \leq L\left(\left\|x_C(t)-x_C(t+\delta t)\right\|+\delta t\right)
\end{equation}

where $L$ is a constant. Since $x_{\mathrm{C}}(t)$ changes little, the $E_C(t)$ remains relatively stable. On the other hand, for the dissipative component $x_D(t)$ which changes significantly over time, the embedding variance:
\begin{equation}
    \left\|E_D(t)-E_D(t+\delta t)\right\| \geq L^{\prime}\left\|x_D(t)-x_D(t+\delta t)\right\|
\end{equation}

where $L'$ reflects the sensitivity of embeddings to dissipative components. The significant changes in $x_D(t)$ leads to higher variance in $E_D(t)$.

\subsubsection{Conservative and Dissipative Genes}
In the context of our dynamical system, the transformation under study is the change in gene embeddings over age. By analyzing these embeddings, we can classify genes into two categories: \textbf{conservative genes}, which exhibit stability in their embeddings over time, and \textbf{dissipative genes}, which show significant changes.

\textbf{Conservative Genes (\( G_c \))}:
Conservative genes are those whose embeddings remain relatively stable across the aging process. This stability indicates that the underlying biological processes associated with these genes are less affected by age-related changes. Formally, we define the set of conservative genes as: 

\begin{equation}
    G_C = \left\{ g \in G \mid \max_{t \in \mathcal{T}} D_g(t) \leq \delta \right\},
\end{equation}

where \(D_g(t)\) represents the embedding drift of gene \(g\) at time \(t\), defined as:
\begin{equation}
    D_g(t) = \left\| E_{g, t} - E_{g, t_0} \right\|,
\end{equation}

with \(E_{g, t}\) denoting the embedding of gene \(g\) at age \(t\) and \(E_{g, t_0}\) as the baseline embedding (embedding at the early lifespan age groups \(t_0\)). The threshold \(\delta > 0\) is a predefined value, determined through statistical analysis, such as a percentile of the \(D_g(t)\) distribution across all genes. These genes show minimal divergence from their baseline embedding, reflecting biological stability and potentially fundamental roles in maintaining homeostasis.

\textbf{Dissipative Genes (\( G_D \))}:
Dissipative genes, in contrast, are those whose embeddings exhibit significant drift over time. They capture processes that are highly dynamic and responsive to age-related changes. The set of dissipative genes is defined as:

\begin{equation}
    G_D = G \setminus G_C
\end{equation}

where \(G\) is the complete set of genes. This classification ensures that any gene not meeting the stability criterion for \(G_C\) is categorized as dissipative.

To quantify the variability of each gene's embeddings over the aging process, we compute the variance of embeddings across all sampled time points:

\begin{equation}
    \sigma_g^2 = \frac{1}{N} \sum_{i=1}^N \left\| E_{g, t_i} - \bar{E}_g \right\|^2,
\end{equation}

where \(N\) is the total number of time points sampled, and \(\bar{E}_g\) is the mean embedding of gene \(g\), calculated as:

\begin{equation}
    \bar{E}_g = \frac{1}{N} \sum_{i=1}^N E_{g, t_i}.
\end{equation}
This variance metric provides an additional layer of insight into the dynamical behavior of genes, allowing us to rank genes based on their temporal stability or variability. Conservative genes will exhibit low variance, while dissipative genes will show high variance, highlighting their active role in dynamic processes. This decomposition into conservative and dissipative genes has some implications for understanding aging-related processes that we will discuss in the result section. 

\subsubsection{Entropy as Metric of Aging}\label{method:entropy}
Entropy metrics offer a complementary perspective on the dynamics of aging by quantifying the uncertainty and complexity in the system. In our model, we analyze entropy changes to understand how the predictability of certain tokens, such as tissue or cell type, evolves over time. These metrics provide insights into the system's variability and stability during the aging process.

To calculate entropy, we focus on the conditional entropy of each masked token, where the model predicts the masked token's value based on the surrounding context. For instance, a masked token might correspond to a specific tissue or cell type. Using the model's predictions, we extract the probability distribution over all possible values (output of the softmax function) and compute the entropy using Shannon entropy. Mathematically, the entropy for a token $x$ is given by:
\begin{equation}
    H(x) = - \sum_{i=1}^k P(x = x_i \mid \text{context}) \log P(x = x_i \mid \text{context}),
\end{equation}

where \(P(x = x_i \mid \text{context})\) is the predicted probability of token \(x\) taking the value \(x_i\), given the surrounding context,\(k\) is size of the dictionary including all tokens, \(H(t)\) is entropy and measures the uncertainty associated with the prediction of token \(x\).


By calculating the entropy for each token at different time points, we can track changes in uncertainty and identify patterns associated with aging. Specifically, tokens with low entropy indicate high confidence in predictions, suggesting stability and consistency in the corresponding biological processes. On the contrary, tokens with high entropy reflect greater uncertainty, potentially pointing to dynamic or less predictable biological processes.

\section{Data Availability}
We incorporated scRNA-seq data from the cellxgene dataset \cite{megill2021cellxgene}, all available publically from their platform. 

\section{Code Availability}
The training script for the model, tokenizer and the code for preprocessing data and inference have been previously published in \cite{khodaee2024multimodal}.

\section{Acknowledgement}
The work was supported in part by a grant from the National Institutes of Health R01 HL161069 awarded to ERE.

\end{document}